\documentclass[apsrev4-1,twocolumn,pra,showpacs,floatfix,superscriptaddress,
citeautoscript]{revtex4-1}
\usepackage{epsf,color,colordvi}
\usepackage{graphicx}
\usepackage{amsmath}
\usepackage{times}
\usepackage{subfigure}
\usepackage{pstricks}
\usepackage{ulem}  
\usepackage{placeins}

\begin{document}
\author{R. K. Bhaduri}
\affiliation{Department of Physics and Astronomy, 
McMaster University, Hamilton, Ontario \L8S 4M1, Canada}
\title{On the higher virial coefficients of a unitary Fermi gas} 
\author{W. van Dijk}\affiliation{Department of Physics and Astronomy, 
McMaster University, Hamilton, Ontario \L8S 4M1, Canada}
\affiliation{Physics Department, Redeemer University College, 
Ancaster, Ontario \ L9K 1J4, Canada}
\author {M. V. N. Murthy}
\affiliation{Institute of Mathematical Sciences, 
Chennai 600113, India}

\begin{abstract} 
Third and higher order quantum
virial coefficients require the solution of the corresponding quantum
many-body problem. Nevertheless, in an earlier paper ( Phys. Rev. Lett. {\bf 108}, 260402 (2012) ) we proposed that the 
  higher-order cluster integrals of a dilute unitary fermionic gas may be
  approximated in terms of the two-body cluster, together with an appropriate  suppression factor. Although not exact, this ansatz gave a fair agreement up to fugacity $z\approx 6$ with the experimentally obtained equation of state. The objective of the present note is to give some physical arguments in favor of this ansatz. 

\end{abstract}


\date{\today} 

\maketitle

\begin{figure}[htpb]
  \centering
\vspace{\baselineskip}
   \resizebox{3.5in}{!}{\includegraphics[width=.5\textwidth, angle=270]{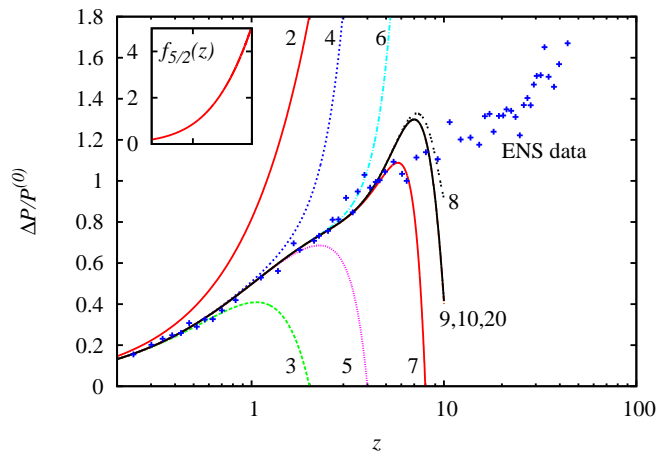}}
   \caption{(Color online) Incremental pressure as a function of fugacity.  The numbers labelling the curves indicate the maximum $l$-value that is included in the sum of Eq.~(\ref{dif}).  The curves with $l=9,10,20$ show no discernible difference.  The insert has a logarithmic horizontal scale indicating a range of $z$ values from 0.2 to 10.  The tic mark corresponds to $z=1$.  The experimental data is taken from Ref.~\cite{1}.}
\label{fig:3}
\end{figure}

Experimentally, it is feasible to adjust the interatomic interaction
in a gas using Feshbach resonance~\cite{ino,ph}. When this is adjusted such that the two atoms are just shy of binding (scattering length $\rightarrow \pm \infty$), the gas is called unitary. In recent times, there has been considerable experimental activity on obtaining the thermodynamic properties of a unitary fermionic gas~\cite{1,2,hori,ku}. It was proposed long back that the equation of state (EOS) of a unitary gas is universal~\cite{ho}, in the sense that when the thermodynamic variables are appropriately scaled, the EOS of  different atomic gases obey the same universal curve. This has been verified experimentally, and has given fresh impetus to the theoretical understanding of such a
gas~\cite{3,3p,Van}. From a theoretical point of view, the EOS of
  a gas may be obtained from a (quantum) virial expansion of the
  grand potential in powers of the fugacity $z$. However, the virial
  coefficient of the $l^{th}$ order term requires a solution of the
  quantum $l$-body problem -- a formidable task for $l>2$.   
In a recent paper~\cite{bvm}, an ansatz was introduced for the higher virial coefficients (for $l \geq 3$ ) of a unitary dilute fermionic gas, that
stated that these higher virial coefficients at high temperatures may
approximately be obtained  by the second virial coefficient multiplied
by an appropriate suppression factor. This is not possible away from unitarity since the clusters of different orders have different temperature dependence~\cite{note}. 
This ansatz for the interaction part of the virial coefficients could fit the 
experimental data of an untrapped gas up to about $z=6$ (see
Fig.~\ref{fig:3}).  The rationale for assuming that it is the two-body cluster  integral that matters in determining the EOS of a unitary gas comes from the realization~\cite{castin} that it is still a very dilute system, with particles primarily undergoing binary scattering. 
 A bigger cluster of particles may be looked upon as mainly a collection of nonoverlapping two-body clusters, provided $z$ is not too large. 
Note, however, that $\Delta b_2=1/\sqrt{2}$ is only for two atoms 
with antiparallel spins, where as in a large cluster there are pairs
with antiparallel as well as parallel spins. In the latter, the atoms
do not interact. In our method, this overcounting is taken care of (in
an average sense) by the suppression factor.  This is only possible  
at unitarity, where all the cluster integrals are taken to be  
 temperature independent. 

Following the notation of Ref.~\cite{bvm}, the grand potential is defined as $\Omega=-\tau \ln {\cal Z}$, where ${\cal Z}(\beta, z) $ is the grand
partition function, $\beta=\tau^{-1}=1/(k_B T)$ is the inverse temperature, and $z=\exp (\beta \mu)$ is the fugacity.  The part of the grand potential coming from the interaction between the atoms may be expanded in a power series of $z$  and written as 
\begin{equation}\label{eq:3}
\Omega -\Omega^{(0)} = -\tau Z_1(\beta)\sum_{l=2}^\infty (\Delta b_l)z^l~.
\end{equation}
The grand potential of the ideal Fermi gas is denoted by
$\Omega^{(0)}$, and is given by $\Omega^{(0)} = -\tau Z_1(\beta) f_{5/2}(z)$, where $f_{\nu}(z)$ is the usual Fermi integral~\cite{pathria}
\begin{equation}
f_{\nu}(z)=\frac{1}{\Gamma (\nu)} \int_0^{\infty} dy \frac{y^{(\nu-1)}}{1+z^{-1}e^y}~.
\end{equation} 
In Eq.~(\ref{eq:3}),  $Z_1(\beta)$ is the one-body partition function, and
$\Delta b_l$ is the $l$-particle  interaction part of the cluster integral.
For an untrapped gas in volume $V$, we have $Z_1(\beta)=2
(V/\lambda^3)$, where spin degeneracy of $2$ is included and 
$\lambda=(2\pi\hbar^2\beta/m)^{1/2}$ is the thermal wave length.  
For a unitary gas, the interaction part of the cluster integrals $\Delta b_l$'s are temperature independent in the high temperature limit. Even though the virial expansion (\ref{eq:3}) for the interaction part converged well even for large $z$, such was not the case for the statistical part $\Omega^{(0)}$. Therefore its exact form by computing the Fermi integral was used, rather than its fugacity expansion.      
From Eq.~(\ref{eq:3}), we see that the interaction part of the grand
  potential requires a knowledge of $\Delta b_l$'s. We proceed to
  obtain these assuming that its major contribution is coming from
  two-body physics.  
It was assumed in Ref.~\cite{bvm} that at unitarity, $\Delta b_l$ could be obtained from $\Delta b_2$ by applying an appropriate suppression factor. We emphasize that the temperature-independent $\Delta b_l$'s at unitarity were obtained only when the quantum expressions were taken to the high temperature limit. There is some justification, then, in doing a  semiclassical analysis by imposing the Pauli principle to the interaction bonds. This has a huge effect on  the counting of bonds. For example, consider the three-body problem. On applying the Pauli
principle, we see that the linked cluster triangle diagram linking all
three particles with interaction bonds is not allowed. This is because two spin-up identical atoms cannot interact at zero-range in the relative 
s-state. 

\begin{figure}[hb]
{
{
$\begin{array}{cc}
\subfigure[]{
\resizebox{0.4\linewidth}{!}{\includegraphics[width=.5\textwidth, angle=0]{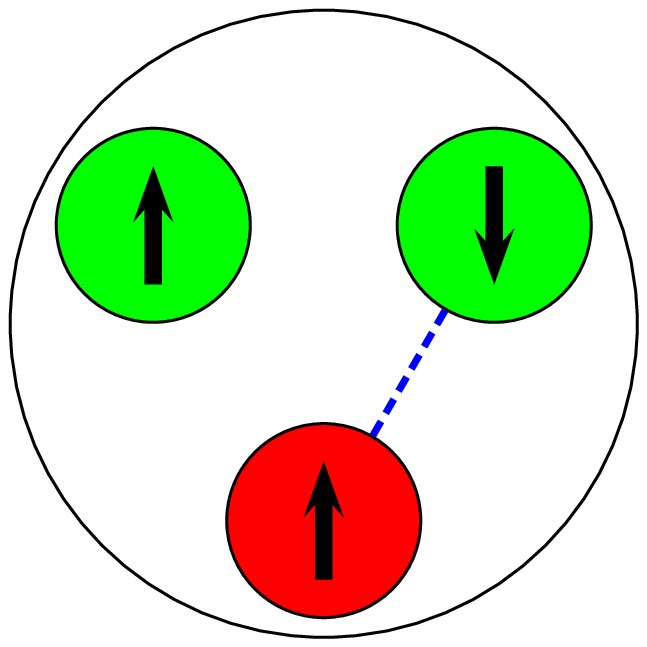}
}
\label{fig:1a} }&
\subfigure[]{
\resizebox{0.4\linewidth}{!}{\includegraphics[width=.5\textwidth, angle=0]{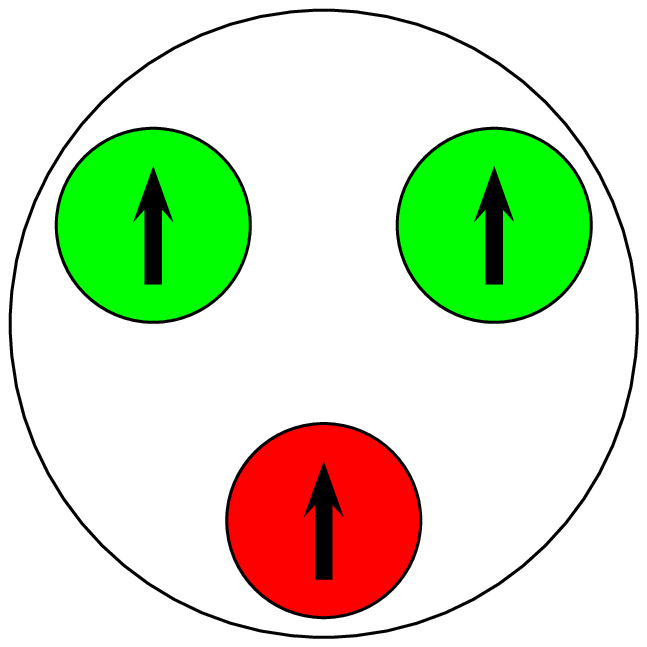}
}
\label{fig:1b}
} \\
\subfigure[]{
\resizebox{0.4\linewidth}{!}{\includegraphics[width=.5\textwidth, angle=0]{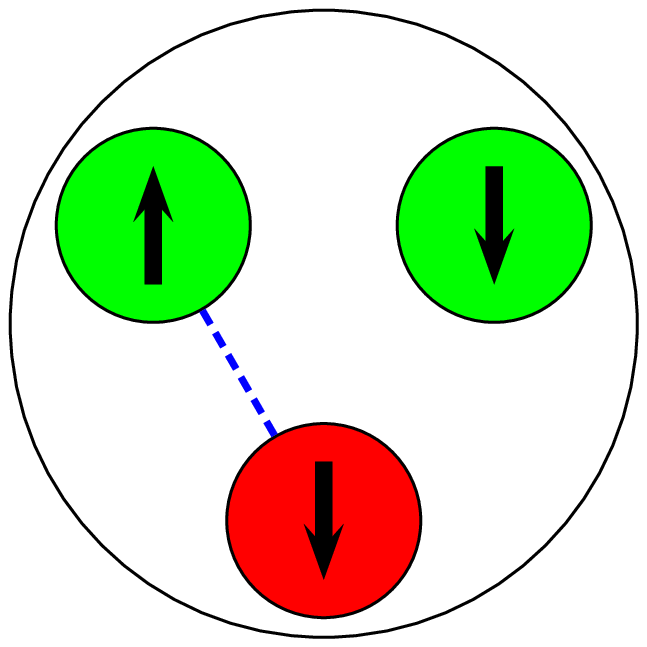}
}
\label{fig:1c}
} &
\subfigure[]{
\resizebox{0.4\linewidth}{!}{\includegraphics[width=.5\textwidth, angle=0]{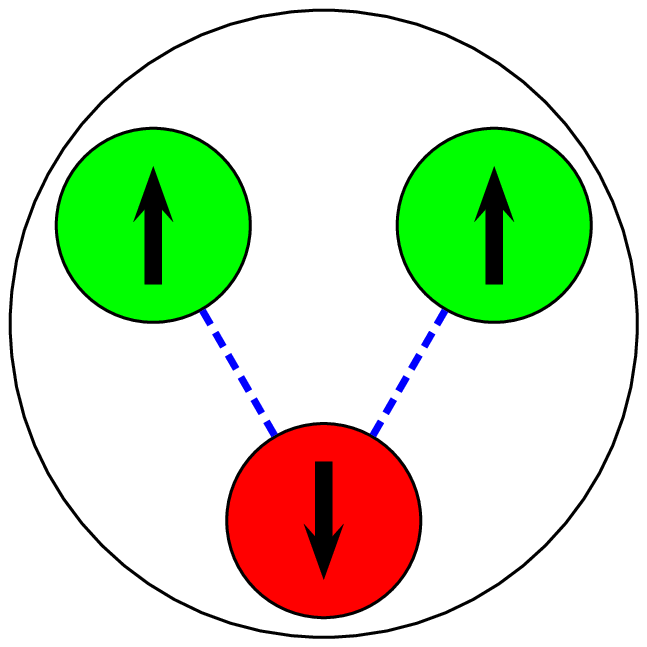}
}
\label{fig:1d}
} \\
\end{array}$
}
\caption{(Color online) Interactions in three-particle clusters.  Effective
  two-particle interactions between the test particle and the pair are indicated by the dashed line connecting the particles.}
\label{fig:1}
}
\end{figure}

\begin{figure}[t]
\resizebox{1.0\linewidth}{!}{\includegraphics[width=.5\textwidth, angle=0]{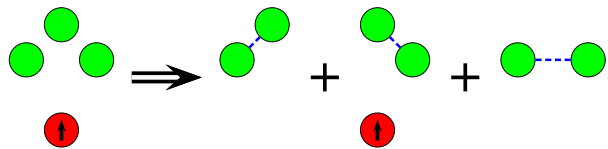}
}
\caption{(Color online) Four-particle cluster.  A test particle effectively sees three pairs.}
\label{fig:2}
\end{figure}

To understand the suppression factor due to Pauli principle,  consider a cluster with $l$ fermionic atoms. Choose any one of them as a test particle, interacting pairwise with the fermions in the remaining $(l-1)$-particle cluster. Our objective is to examine how the two-body bonds involving the test particle with the rest may get suppressed due to the Pauli blocking.  Let ${\cal N}_{(l-1)}$ denote the number of two-body pairs in a cluster with $(l-1)$ fermions. To illustrate with the simplest example, let $l=3$. For this case ${\cal N}_{(l-1)}=(l-1)(l-2)/2=1$, and the test particle sees only one pair, as shown in Fig.~\ref{fig:1}.
 In Fig.~\ref{fig:1a},  we assume that the test particle has spin up, and the pair consists of one spin-up and the other spin-down particle. 
Since the interaction is a zero-range (s-state) potential, the test particle interacts with only one of the two possible bonds. Thus there is a suppression factor of 2 due to the Pauli principle. In Fig.~\ref{fig:1b}, the test particle is still spin-up, but the two particles in the pair are both spin-up. In this situation, the test particle cannot interact with either of the particles in the pair. In Figs.~\ref{fig:1c} and \ref{fig:1d}, the test particle is spin down. We see that in 1(d) the test particle now interacts with both the particles in the pair, thus the average suppression of (b) and (d) is  still a factor of 2.  Next, we examine the $l=4$ case shown in
Fig.~\ref{fig:2}. Now a test particle within this cluster sees a
three-particle cluster, which may be looked upon, in the dilute gas,
as three pairs. For every pair that the test particle sees, there is a suppression factor of 2, so that the net suppression factor is $2^3$. Extending this line of reasoning to higher order   clusters, we apply a suppression factor of   $2^{{\cal N}_{(l-1)}}$ in the number of two-body bonds within a  $l$-body cluster.  
What about the sign of $\Delta b_l$ ?  
Note that for the ideal Fermi gas, the statistical virial clusters
$b_l^{(0)}$'s have an alternating factor $(-1)^{(l+1)}$. Since the
Pauli principle gives an effective repulsive effect, in contrast to the
attractive potential at unitarity, we expect the $\Delta b_l$'s to be
of opposite signs to $b_l^{(0)}$'s. 

Using these reasonings, we write down the equation for $\Delta b_l$'s  
that was given in~\cite{bvm} :   
 \begin{equation}\label{eq:4}
\Delta b_l = (-)^l \dfrac{(\Delta b_2)}{2^{{\cal N}_{(l-1)}}},
\ \ l\geq 2.
\end{equation}
In the above, as stated earlier, ${\cal N}_{(l-1)}=(l-1)(l-2)/2$ 
is the number of pairs in a cluster with $(l-1)$ fermions.  
For $l=2, \ {\cal  N}_1=0$, and Eq.~(\ref{eq:4}) is an identity. 
Since $\Delta b_2=\frac{1}{\sqrt{2}}$ is known analytically~\cite{HoMu},  
all the higher virial coefficients can be found using our Eq.~(\ref{eq:4}). 
 The third virial coefficient has been calculated very
 accurately~\cite{liu,rakshit} up to 12 decimal figures to be 
 $-0.3551...$. Our formula gives $\Delta b_3=-\frac{1}{2\sqrt{2}}=-0.3536..$.
For $l=4$, we get $\Delta b_4=\frac{1}{8 \sqrt{2}}=0.088..$, to be
compared with the value based on measurements, $0.096\pm 0.015$~\cite{1}. 

The alternating signs in Eq.~(\ref{eq:4}) are borne out by  experimental data, up to at least $l=8$.  (See Fig.~\ref{fig:3}.)  Using Eq.~(\ref{eq:3}), we get 
\begin{equation}
\frac{\Delta P}{P^{(0)}}=\frac{\sum_{l=2}^{\infty} (\Delta b_l)  z^l}{f_{5/2}(z)}~,
\label{dif}
\end{equation} 
where $P$ is the pressure of the interacting gas, and $P^{(0) }$ the 
pressure of an ideal Fermi gas at the same value of $z$, and in the
same volume. In the above equation, $\Delta P=P-P^{(0)}$ shows that a
positive $\Delta b_l$ increases the pressure from its ideal value, where as a negative $\Delta b_l$ decreases it. This is seen clearly in Fig.~\ref{fig:3}, where the incremental contribution to the pressure is shown by taking the upper limit in the summation of $l$ at $l=2,3,4..$ etc. In the curve labelled 3, for example, only the contributions from $\Delta b_2$ and $\Delta b_3$ are included. The alternating signs $(-1)^l$ ensure that the virial series for the EOS follows the experimental points closely.  It is evident from Fig.~\ref{fig:3} that the series converges since terms involving $l>9$ do not change the sum.    
  
We note that in general the $l$-body quantum problem has to be solved in order to obtain the cluster integral $\Delta b_l$. We have argued,
however, that in the high temperature limit of a dilute unitary gas,
the main contribution to the $l$-body cluster comes from two-body physics.  It is essential to go to the high temperature limit where the virial coefficients are temperature independent, and semiclassical arguments may be made. The qualitative physical arguments   
led us to propose Eq.~(\ref{eq:4}). Since it is able to match the
experimental data well, this presentation may encourage others to do  
a more quantitative derivation of the equation.


\begin{thebibliography}{99}
\bibitem{note} In a many-body classical system with only hard-core
  pairwise interaction, the higher virial coefficients are
  proportional to the powers of the second virial coefficient, despite
  having temperature dependence; see Pathria (next reference).
\end{thebibliography}
\end{document}